\documentclass[pra,twocolumn, showpacs,showkeys,secnumarabic,amssymb,amsmath,unsortedaddress]{revtex4-1}
\usepackage{amssymb,amsmath}
\usepackage{color}
\usepackage{graphics}
\usepackage{graphicx}
\usepackage{enumitem}
\usepackage{bm}
\begin{document}

\title{Quantum coherence of planar spin models with Dzyaloshinsky-Moriya interaction}

\author{Chandrashekar Radhakrishnan}
\affiliation{New York University Shanghai, 1555 Century Avenue, Pudong, Shanghai 200122, China}
\affiliation{NYU-ECNU Institute of Physics at NYU Shanghai, 1555 Century Avenue, Pudong, Shanghai 200122, China}

\author{Igor Ermakov}
\affiliation{New York University Shanghai, 1555 Century Avenue, Pudong, Shanghai 200122, China}
\affiliation{ITMO University, Kronverkskiy 49, 197101, St.Petersburg, Russia}
\affiliation{Saint Petersburg State University, University Embankment, 7-9, St Petersburg, Russia}

\author{Tim Byrnes}
\affiliation{State Key Laboratory of Precision Spectroscopy, School of Physical and Material Sciences, East China Normal University,
Shanghai 200062, China}
\affiliation{New York University Shanghai, 1555 Century Avenue, Pudong, Shanghai 200122, China}
\affiliation{NYU-ECNU Institute of Physics at NYU Shanghai, 1555 Century Avenue, Pudong, Shanghai 200122, China}
\affiliation{National Institute of Informatics, 2-1-2 Hitotsubashi, Chiyoda-ku, Tokyo 101-8430, Japan}
\affiliation{Department of Physics, New York University, New York, NY 100003, USA}

\begin{abstract}
The quantum coherence of one dimensional planar spin models with the Dzyaloshinsky-Moriya interaction is investigated.  
The anisotropic XY model, the isotropic XX model and the transverse field model are studied in the large $N$-limit using the 
two qubit reduced density matrices and the two point correlation functions.  From our investigations we find that the 
coherence as measured using the Jensen-Shannon divergence can be used to detect the quantum phase transitions and the quantum 
critical points.  The derivative of coherence shows non-analytic behavior at the critical points 
leading to the conclusions that these transitions are of second order.  Further we show that the presence of the Dzyaloshinsky-Moriya 
coupling suppresses the phase transition due to the residual ferromagnetism which is caused by spin canting.  
\end{abstract}


\maketitle

%
%
%
\section{Introduction}
\label{Intro}
Quantum systems are correlated in ways which are quite different from classical systems.  These quantum correlations manifest 
themselves in many physical forms. One of the well known and widely studied quantum correlation is entanglement 
\cite{einstein1935can,peres1996separability,horodecki1997separability,horodecki2009quantum}.  But entanglement
is not the only form of quantum correlations, since it accounts only for non-local quantum correlations.  There are several 
other quantities such as quantum discord \cite{henderson2001classical,ollivier2001quantum} and quantum dissonance \cite{modi2010unified} to investigate quantum correlations which are not accounted by 
non-local quantum correlations.  For any practical application a comprehensive understanding of all the quantum correlations is 
essential since quantumness can be exhibited by separable (non-entangled) states.  

In the past few decades the application of quantum information theory to condensed matter physics has been on the rise.  Most 
of these applications involved studying quantum critical phenomenon using entanglement theories \cite{osborne2002entanglement,latorre2009short}.
In classical systems thermal fluctuations causes a phase transition which can be detected from the variation of thermodynamic 
parameters like entropy or free energy.  Quantum phase transitions arise due to quantum fluctuations originating from the 
Heisenberg uncertainity principle \cite{sachdev2007quantum}.  
A wide range of systems exhibit quantum phase transitions and their characterization 
always involves the study of quantum correlations in the system.  This is because quantum correlations can capture the 
quantum fluctuations in a system.  This line of thinking has led to characterization of quantum phase transitions by examining the entanglement 
in the system.  In spin systems most often the pairwise entanglement is used \cite{osborne2002entanglement,wu2004quantum}.  Numerous studies have indicated that 
measures of entanglement such as concurrence \cite{bennett1996mixed,hill1997entanglement,wootters1998entanglement},
quantum entropy  \cite{osborne2002entanglement}, and quantum fidelity  \cite{zanardi2006ground} can give information about the quantum phase transition 
\cite{osborne2002entanglement,osterloh2002scaling,gu2003entanglement,gu2004entanglement} and  quantum critical points 
\cite{vidal2003entanglement,vidal2004entanglement}. 

A common feature of these measures is that they quantify bipartite entanglement, which is known to be a short ranged quantity 
\cite{osborne2002entanglement,ccakmak2015factorization}.   However, phase transitions are the result of long-ranged correlations and have been observed in some cases \cite{giampaolo2014genuine} to be insufficient to characterize quantum phase transitions.  This has led to the study of alternative  quantities which give a better understanding of quantum critical systems, such as  multipartite entanglement \cite{de2006multipartite,de2006genuine,giampaolo2013genuine,hofmann2014scaling} and bipartite correlation measures such as quantum discord 
\cite{ollivier2001quantum,henderson2001classical} 
and quantum coherence \cite{baumgratz2014quantifying}, which are potentially long-ranged.   One of the drawbacks of multipartite entanglement is that it is computationally expensive to calculate since it involves diagonalizing a $2^n \times 2^n$ matrix to estimate the $n$-partite entanglement. On the other hand, bipartite quantum correlation approaches have the advantage that it can be calculated from a $4 \times 4$ matrix (for spin-$1/2$), which is computationally more efficient. Additionally, it is not a requirement that 
quantum phase transitions are accompanied by the presence of entanglement, for example in systems at non-zero temperature \cite{werlang2010quantum}.  This motivates other quantities such as quantum discord \cite{dillenschneider2008quantum,sarandy2009classical,werlang2010quantum,werlang2010thermal,chen2010quantum} which detect quantum correlations, and quantum coherence as we investigate in this paper.

Recently, there has been a resurgence of interest in quantum coherence, following from the seminal work of Baumgratz, Cramer, and Plenio where a rigorous framework for quantifying it
was established in the language of quantum information theory \cite{baumgratz2014quantifying}.  Though earlier schemes in \cite{glauber1963coherent,sudarshan1963equivalence} 
gave procedures for detecting coherence, it was never quantified in a rigorous sense. 
Based on this new method of quantifying coherence several developments were made towards understanding coherence 
from the point of view of information theory
 \cite{yao2015quantum,yadin2015quantum,ma2016converting,yadin2016general,bromley2015frozen,du2015conditions,pires2015geometric,
cheng2015complementarity,streltsov2015measuring} and in using it as a resource in quantum information theory 
\cite{winter2016operational,brandao2015reversible,chitambar2016critical,chitambar2015relating,del2015resource}.  
Consequently, many investigations of thermodynamic processes 
\cite{lostaglio2015description,narasimhachar2015low,lostaglio2015quantum} and experimentally feasible systems 
\cite{man2015cavity,wang2016gaussian,zheng2016detecting,opanchuk2016quantifying}
 were carried out. 
Ideally the coherence of given state is measured as its distance to the closest incoherent state and all these 
measures of coherence \cite{baumgratz2014quantifying,shao2015fidelity,rana2016trace,
zhang2016quantifying,xu2016quantifying,adesso2016measures,yao2016frobenius,napoli2016robustness}
may be broadly classified into either geometric or the entropic class of measures each
with its own advantages. We previously introduced a new measure, based on the quantum version of the Jensen-Shannon divergence,  which has both entropic and geometric features.  This feature is convenient as it allows for an analysis of the inter-qubit correlations (``intrinsic coherence'') and the intra-qubit coherence (``local coherence'').  A scheme to measure each of these coherences was also proposed along with 
the relationship between them.

Investigations on quantum coherence in complex many-body systems are limited to a few works \cite{karpat2014quantum,li2016quantum,cheng2016finite,ccakmak2015factorization}.  In particular, the Jensen-Shannon
divergence has not been used to investigate quantum phase transitions.  For other coherence measures, several studies were carried out in spin chain models \cite{malvezzi2016quantum} 
in particular the XY model \cite{karpat2014quantum,li2016quantum,cheng2016finite,ccakmak2015factorization} with a view to understand the
 role played by quantum coherence in quantum phase transitions.  These studies 
 generally show that coherence can be used to detect important 
features such as the critical points, but due to the few models and measures it is still difficult to say how robust quantum coherence is as 
a detector of quantum phase transitions.

Further all these models consider only the symmetric spin-spin interaction.  
The antisymmetric Dzyaloshinsky-Moriya interaction was introduced
through the works of Dzyaloshinsky \cite{dzyaloshinsky1958thermodynamic} and Moriya \cite{moriya1960anisotropic} to explain the weak ferromagnetic behavior in the antiferromagnetic state of
certain systems like $\alpha$-$\text{Fe}_{2}\text{O}_{3}$, $\text{MnCO}_{3}$ and $\text{CoCO}_{3}$.  
It was shown that the spin-lattice and dipole 
interactions causes the spins to be tilted by a small angle with respect to each other rather than being exactly co-parallel
to each other.  This phenomenon is called spin canting and it gives rise to a net effective magnetism in the system.  Using 
crystal symmetry arguments Moriya showed in Ref. \cite{moriya1960anisotropic} that the DM interaction 
can be explained through an antisymmetric exchange 
interaction term.  Thus an investigation of this model can help us to 
understand the use of quantum coherence to study quantum phase transition 
in a wider class of models whose behaviour cannot be accounted for by using only the 
spin-spin interaction.

In this paper, we consider an XY spin model with both the symmetric spin-spin interaction and the antisymmetric Dzyaloshinsky-Moriya (DM) interaction.  This is an analytically solvable model with well-known physical properties hence it is a suitable testing ground for the coherence approach to detecting quantum phase transitions using the QJSD.  We investigate the coherence of this model in the thermodynamic limit at $T=0$ and its connection to quantum phase transitions.
 This paper is structured as follows.   An overview of the XY model with DM interaction and the two point correlation functions is reviewed in Sec. \ref{xydm}.  The quantum coherence of planar spin models with DM interaction and its connection 
to quantum phase transition is studied in Sec. \ref{qcqpt}. In this section we examine several cases where the various parameters in the model are varied.   A summary of our conclusions is presented in 
the Sec. \ref{conclusions}.

%
%
%
\section{XY model with DM interaction}
\label{xydm}

A one-dimensional Heisenberg XY chain with spin-orbit coupling with periodic boundary conditions in the presence
of an external field is described by the following Hamiltonian: 
\begin{eqnarray}
H = &\sum_{i=1}^{N}& J\Big[(1+\gamma) \sigma_{i}^{x} \sigma_{i+1}^{x} 
                    + (1-\gamma) \sigma_{i}^{y} \sigma_{i+1}^{y} \nonumber \\
  & &               + D (\sigma_{i}^{x} \sigma_{i+1}^{y} - \sigma_{i}^{y} \sigma_{i+1}^{x}) \Big] 
                    - \sigma_{i}^{z},
\label{xydmhamiltonian}                     
\end{eqnarray}
where $-1 \leq \gamma \leq 1$ is the anisotropy parameter.  Here we have assumed that the spin-orbit coupling and 
the external field to be along the $z$-direction. The limits $\gamma=0$ and $\gamma = \pm 1$ corresponds to the 
isotropic XX model and the Ising model respectively.  The system is antiferromagnetic when $J > 0$ and for $J<0$ it is 
ferrromagnetic.  The factor $D$ describes the strength of the antisymmetric DM interaction. 
In the current work we measure the quantum coherence of the system described by the Hamiltonian in 
(\ref{xydmhamiltonian}) in the $N \rightarrow \infty$ (thermodynamic) limit and the zero temperature limit.  

The XY model was solved exactly by Lieb, Schultz and Mattis \cite{lieb1961two} for the case of a zero magnetic field by Niemeijer for a
finite external field \cite{niemeijer1967some,niemeijer1968some}.  The correlation functions of the model were introduced in 
\cite{lieb1961two,niemeijer1967some,niemeijer1968some} and extensive investigation of these
correlation functions were provided in \cite{barouch1970statistical,barouch1971statistical}.  
The conventional approach to solve the XY-model involves expressing the spin
operators $\sigma^{\alpha}_{i}$ in terms of the creation/annihilation operators of spinless fermions through the 
Jordan-Wigner transformation \cite{giamarchi2004quantum}.  Under such transformation the Hamiltonian of the XY-model 
transforms into free fermion
Hamiltonian, which enables us to derive an exact expression for its corelation functions.  It turns out that the 
Hamiltonian (\ref{xydmhamiltonian}) being mapped into fermions is also free, hence the XY-model with DM interaction 
can be solved exactly as well.

Using the Jordan-Wigner transformation \cite{giamarchi2004quantum} as in 
\cite{barouch1970statistical,barouch1971statistical}, the exact expressions for an energy spectrum of the model and the 
$N$-point correlation functions can be derived \cite{liu2011quantum}.  
The quantum correlation between two spins at 
sites $i$ and $j$ can be found from their collective state.  The two site reduced density matrix $\rho(i,j)$ contains 
information about the collective state.  This reduced density matrix can be obtained from the total density matrix using the 
relation $\rho(i,j) = {\rm Tr}_{\overline{ij}} \; \rho$, i.e., by tracing over all the qubits other than those corresponding to 
the $i$ and the $j$ sites. Below we give the two qubit reduced density matrix in the computational basis \cite{liu2011quantum}: 
\begin{equation}
\rho(i,i+r) = 
\begin{pmatrix}
u_{+} & 0 & 0 & x_{-} \\
0 & w & x_{+} & 0 \\
0 & x_{+} & w & 0 \\
x_{-} & 0 & 0 & u_{-} 
\end{pmatrix},
\label{reduceddm}
\end{equation}
The $U(1)$ invariance and the symmetries of the Hamiltonian (\ref{xydmhamiltonian}) gives rise to the form 
of the two qubit density matrix given in (\ref{reduceddm}).  For the sake of completeness we give a list of the symmetries below: 
 The translational invariance of the Hamiltonian ensures that the density matrix verifies $\rho(i,j) = \rho(i, i+r)$ implying 
that the density matrix is a function of the number of lattice sites between them and independent of the actual location 
of the spins.  Since the Hamiltonian is real the density matrix is also real $\rho(i,j) = \rho(i,j)^{*}$, 
further due to the reflection symmetry we also have $\rho(i,j) = \rho(j,i)$. The elements of the density matrix in 
(\ref{reduceddm}) in terms of the two point correlation functions are
\begin{eqnarray}
u_{\pm} &=& \frac{1}{4} \pm \frac{\langle \sigma_{i}^{z} \rangle}{2} 
          + \frac{\langle \sigma_{i}^{z} \sigma_{i+r}^{z} \rangle}{4}, \nonumber \\
w &=& \frac{1 - \langle \sigma_{i}^{z} \sigma_{i+r}^{z} \rangle}{4}, \nonumber\\
x_{\pm} &=& \frac{\langle \sigma_{i}^{x} \sigma_{i+r}^{x} \rangle \pm \langle \sigma_{i}^{y} \sigma_{i+r}^{y} \rangle}{4}.
\label{dmelements}
\end{eqnarray}
The magnetization of the XY model with DM interaction in the presence of an external field is
\begin{equation}
\langle \sigma_{i}^{z} \rangle  = -\frac{1}{\pi} \; \int_{0}^{\pi} {\rm d} \phi 
                             \frac{[J(\cos \phi - 2 D \sin \phi) -1]}{\Delta},
\label{magnetizationxy}                             
\end{equation}
where 
\begin{equation}
\Delta = \sqrt{[J(\cos \phi - 2D \sin \phi)-1]^{2} + J^{2} \gamma^{2} \sin^{2} \phi}.
\label{deltadefinition}
\end{equation}
The two point spin-spin correlation functions corresponding to the $x$ and the $y$ axis can 
be computed from the determinant of the Toeplitz matrices and for the $z$ axis it can be 
evaluated from the magnetization.  The mathematical form of these quantities reads:
\begin{eqnarray}
\langle \sigma_{i}^{x} \sigma_{i+r}^{x} \rangle &=& 
\begin{vmatrix}
G_{-1} & G_{-2} & \cdots & G_{-r}\\
G_{0}  & G_{-1} & \cdots & G_{-r+1} \\
\vdots & \vdots & \ddots & \vdots \\
G_{r-2} & G_{r-3} & \cdots & G_{-1} \\
\end{vmatrix},
\label{xxcorrelator}
\\
\langle \sigma_{i}^{y} \sigma_{i+r}^{y} \rangle &=& 
\begin{vmatrix}
G_{1} & G_{0} & \cdots & G_{-r+2}\\
G_{2}  & G_{1} & \cdots & G_{-r+3}\\
\vdots & \vdots & \ddots & \vdots \\
G_{r} & G_{r-1} & \cdots & G_{1} \\
\end{vmatrix},
\label{yycorrelator}
\\
\langle \sigma_{i}^{z} \sigma_{i+r}^{z} \rangle &=& \frac{\langle \sigma^{z} \rangle^{2} -  G_{r} \, G_{-r}}{4},
\label{zzcorrelator}
\end{eqnarray}
where,
\begin{eqnarray}
G_{k} &=&  - \frac{1}{\pi} \int_{0}^{\pi} {\rm d} \phi \frac{2 \cos(\phi k)}{\Delta} \,
           [J (\cos \phi - 2 D \sin \phi) - 1] \nonumber \\ 
      &  &  + \frac{\gamma}{\pi} 
            \int_{0}^{\pi} {\rm d}  \phi \; \frac{2 J \sin (\phi k)}{\Delta} \sin \phi.
\label{greensfunction}            
\end{eqnarray}
Hence the quantum correlations in the XY Model with DM interaction can be found from the knowledge of the two-qubit reduced
density matrix (\ref{reduceddm}) and the two-point correlation functions which are (\ref{xxcorrelator})-(\ref{zzcorrelator}).  
In the following sections we use them to evaluate the quantum coherence.  

\begin{figure*}[ht]
\includegraphics[width=160mm]{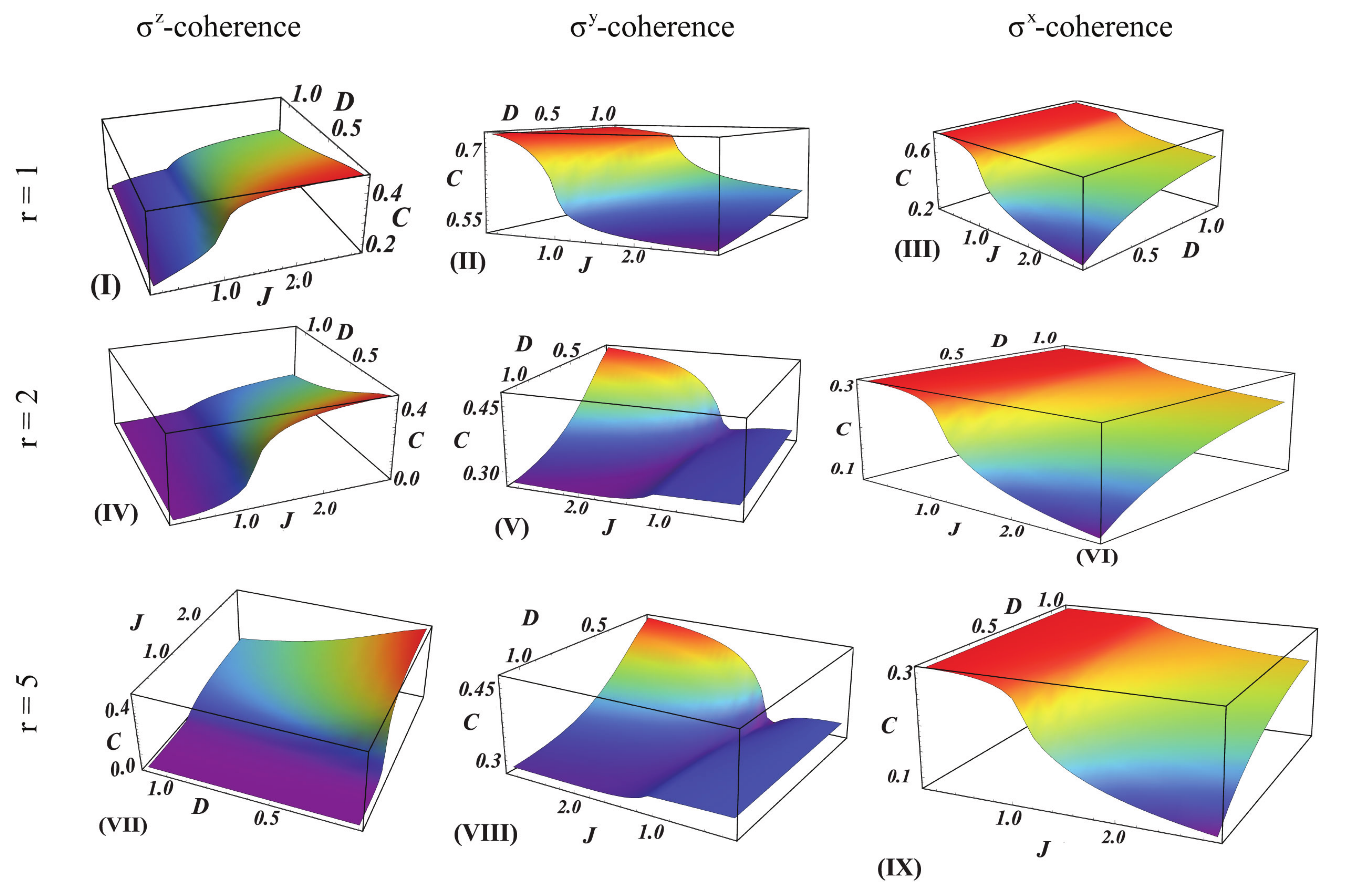}
\caption{The variation of quantum coherence $C$ with respect to the interaction parameters $J$ and 
Dzyaloshinsky-Moriya coupling parameter $D_z$ is discussed for the $XY$ model with anisotropy parameter 
$\gamma = 0.5$. There are nine 3-D plots presented in a grid type fashion where the first row 
$(I,II,III)$, second row $(IV,V,VI)$ and the third row $(VII,VIII,IX)$ correspond to the nearest 
neighbour, the next nearest neighbour and the fifth nearest neighbour.  The first column $(I,IV,VII)$,
the second column $(II, V, VIII)$ and the third column $(III,VI,IX)$ are the quantum coherences in the 
$\sigma^z$, $\sigma^y$ and the $\sigma^x$ bases respectively.}
\label{fig1}
\end{figure*}

%
%
%
\section{Quantum Coherence and Quantum Phase transition}
\label{qcqpt}
\subsection{Quantum Jensen Shannon divergence}

The quantum coherence of the XY model with DM interaction described via the Hamiltonian (\ref{xydmhamiltonian}) is 
examined in the current section.  For quantifying the amount of coherence we use the measure based on the quantum version of the 
Jensen-Shannon divergence which was introduced in \cite{radhakrishnan2016distribution}.  
The Jensen-Shannon divergence \cite{lin1991divergence} is based on the entropic function and 
also obeys the distance properties unlike the relative entropy which does not obey the symmetry axiom of a distance measure. 
If the measure is based on the square root of the Jensen-Shannon divergence it verifies the triangle inequality in addition to 
satisfying the distance axioms, which is advantageous since the measure has metric properties.  For pure states, the metric nature of 
the measure based on the square root of the Jensen-Shannon divergence has been proved analytically for arbitrary number of qubits 
\cite{lamberti2008metric,briet2009properties}.  
In the case of mixed states a numerical proof exists for systems with up to five qubits.  The measure of quantum coherence based 
on the square root of the Jensen-Shannon divergence is 
\begin{equation}
C(\rho) = \sqrt{S\left( \frac{\rho+\rho_d}{2} \right) - \frac{S(\rho)}{2}-\frac{S(\rho_d)}{2}}.
\label{coherencedef}
\end{equation} 
Here, $\rho_{d}$ is the incoherent state corresponding to the density matrix and $S(\rho) =  - {\rm Tr} \rho \log_{2} \rho$ is the 
von Neumann entropy.  

The reduced density matrix of two arbitrary sites given in (\ref{reduceddm}) has been computed in the $\sigma^z$-basis.  From the 
structure of the density matrix in the $\sigma^z$-basis we observe that it is an X-state (a state with only diagonal and antidiagonal
elements).  Meanwhile, the incoherent state corresponding 
to the density matrix is completely diagonal and the matrix $(\rho+\rho_{d})/2$ is also an X-state.  The eigenvalues of the 
density matrix and the matrix $(\rho+\rho_{d})/2$ can be calculated using the expressions:
\begin{eqnarray*}
\lambda_{1,2} &=& \frac{1}{2} [(\rho_{11} + \rho_{44}) \pm \sqrt{(\rho_{11} - \rho_{44})^{2} + 4 |\rho_{14}|^{2}}], \\
\lambda_{3,4} &=& \frac{1}{2} [(\rho_{22} + \rho_{33}) \pm \sqrt{(\rho_{22} - \rho_{33})^{2} + 4 |\rho_{23}|^{2}}],
\end{eqnarray*}
where $\rho_{nm}$ is the element of the density matrix with $n$ and $m$ being the row and the column index.  Since the density matrix
of the incoherent state is completely diagonal the eigenvalues are easy to find.  From the knowledge of the eigenvalues, one can find 
the respective entropies, using which the amount of coherence in the system can be computed using equation (\ref{coherencedef}).

\begin{figure}[ht]
\includegraphics[width=\linewidth]{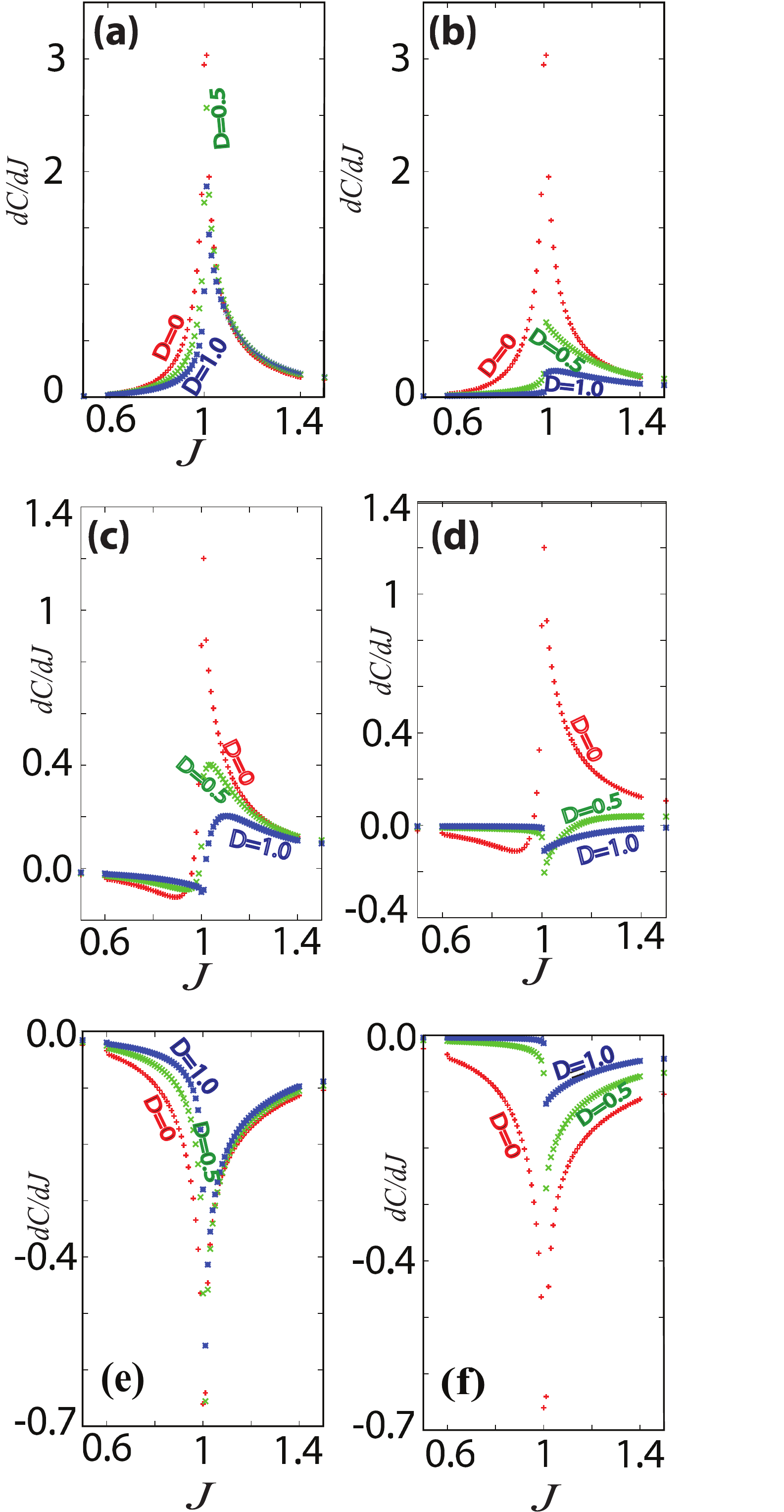}
\caption{The derivative of quantum coherence $dC/dJ$ is plotted as a function of the spin-spin interaction strength $J$
for the value of $\gamma =0.5$. 
The plots, Fig. 2(a) and 2(b) correspond to the derivative of $\sigma^z$-basis coherence for the nearest neighbour and the fifth neighbour
for different $D$. 
Similarly Fig's 2(c) and 2(d) show the derivative of $\sigma^y$-basis coherence for the nearest neighbour and the fifth neighbour for 
various values of the Dzyaloshinsky-Moriya coupling parameter $D$. 
Finally Fig's 2(e) and 2 (f) show the derivative of $\sigma^{x}$-basis coherence for various values of $D$.  
}
\label{fig2}
\end{figure}

Quantum coherence measures usually find the distance to the incoherent state corresponding to the density matrix under investigation. 
If we consider a density matrix $\rho$ in a particular basis $b_{1}$, of a state existing in a $d$-dimensional Hilbert space, the 
incoherent state is defined as a state which is diagonal in this basis.  There are many diagonal states in a particular basis and we 
estimate the distance to the closest incoherent (diagonal) state.  Any state which is diagonal in a basis $b_{1}$ may not be diagonal
in another basis $b_{2}$ and so it will not belong to the set of the incoherent states in the $b_{2}$ basis.  Thus we can observe that 
the set of incoherent states is completely dependent on the basis and so the coherence naturally becomes a basis dependent quantity.  
Our procedure described so far in this section gives us the means to investigate the coherence in the $\sigma^z$-basis since the density matrix
in equation (\ref{reduceddm}) is in the $\sigma^z$-basis.  To compute the coherence in the alternate bases (either $\sigma^x$ or $\sigma^y$)
we make a unitary transformation of the density matrix $\rho$ from the $\sigma^z$-basis to the new basis. The incoherent state corresponding to 
the density matrix in the new basis and the state $(\rho+\rho_{d})/2$ are found.  Using the density matrices in the new basis we compute 
the corresponding entropies and consequently the coherence in the new basis.  The entropy of the density matrix will not change since the 
eigenvalues are invariant under unitary transformation, but the entropies of the matrices $\rho_{d}$ and $(\rho+\rho_{d})/2$ will change
since the matrix $\rho_{d}$ depends on the choice of the basis. In the present work we estimate the coherence along all the 
three orthogonal directions namely $\sigma^x$, $\sigma^y$ and $\sigma^z$ for the nearest neighbour ($r=1$), the next nearest neighbour ($r=2$) and the 
fifth nearest neighbour ($r=5$).  

\subsection{Anisotropic case: $\gamma = 1/2$}

Based on our computations we plot the variation of coherence with respect to the spin-spin interaction parameter $J$ and the 
spin-orbit interaction parameter $D$.  The results are show in Fig. \ref{fig1}.  From the plots we observe 
that in the $\sigma^z$-basis the quantum coherence in the system increases as the spin-spin interaction increases.  But when the DM 
interaction is also varied the change in coherence is much slower compared to the situation when $D=0$.  In the case of $\sigma^y$-basis 
we observe two different behaviors one for the nearest neighbour and the other type of behavior for the next nearest and the 
fifth neighbour.  For the nearest neighbour the quantum coherence in the system decreases as $J$ increases but the fall of coherence
is much slower when $D$ is higher.  In the case of the next nearest and the fifth neighbour, the coherence either increases to a 
maximum value or decreases to zero depending on the value of $D$. The $\sigma^x$-basis coherence decreases as $J$ increases but 
a higher value of DM interaction parameter slows down the rate of decrease.  

From Fig. \ref{fig1} we can observe that there is a significant change in quantum coherence in the region around $J=1$, which is the 
critical point.  A more precise knowledge of the location of the critical point and the order of phase transition can be obtained 
from the derivatives of the quantum coherence with respect to the coupling parameter $J$. 
According to the principles of statistical mechanics, if the first derivative of the energy presents a finite 
discontinuity then we have a first order transition. But if the first derivative is continuous and the second derivative shows 
discontinuity or divergence then the quantum phase transition is of second order.  These observations were related to the 
information theoretic quantities in \cite{wu2004quantum} by connecting them to the reduced density matrix and its derivatives.  
A discontinuity in the first derivative of the energy signifies that there is a discontinuity in 
atleast one of the reduced density matrices. Analogously a divergence
in the derivative of the density matrix arises when there is a discontinuity or divergence in the second derivative of the energy.  
Thus a knowledge of the nonanalyticities of the density matrix and its derivatives with respect to the coupling parameter $J$ 
can gives us information about the order of the phase transition and the location of the critical point. 

Quantum coherence as estimated
using (\ref{coherencedef}) involves the entropies which are functions of the density matrices, hence should give us information 
about the quantum phase transitions.  If the quantum coherence defined using (\ref{coherencedef}) shows discontinuity then it 
implies that there is a first order phase transition and if the first derivative of our coherence measure shows divergence then 
we can conclude it is a second order phase transition.  The results in Fig. \ref{fig1} show that there is no discontinuity in the 
quantum coherence and hence it is not a first order phase transition. 
The derivative with respect to $ J $ however shows a discontinuity, signaling a phase transition.  In Fig. \ref{fig2} we plot $ \frac{dC}{dJ} $ in the three mutually orthogonal $\sigma^{z}$, $\sigma^{y}$ and the $\sigma^{x}$ bases respectively.  The 
discontinuity in the derivative of coherence with respect to $J$ is in the positive and negative axes for the $z$-basis and the 
$x$-basis respectively as given in Figures \ref{fig2}(a)(b)(e)(f).  In the case of the $y$-basis for the 
fifth nearest neighbor (Fig. \ref{fig2}(d)), we find that the discontinuity of the derivative changes its direction from the positive 
to the negative axes with increase in the value of the DM parameter.  Though there is a change in direction of the derivative of coherence 
as $D$ changes, the phase transition for this model can be detected for all values of $D$.  This can be observed from 
Figure \ref{fig1}(VIII) where we notice that for all values of $D$ there is at least one point where the coherence is not differentiable.

\begin{figure*}[ht]
\centering
\includegraphics[width=\linewidth]{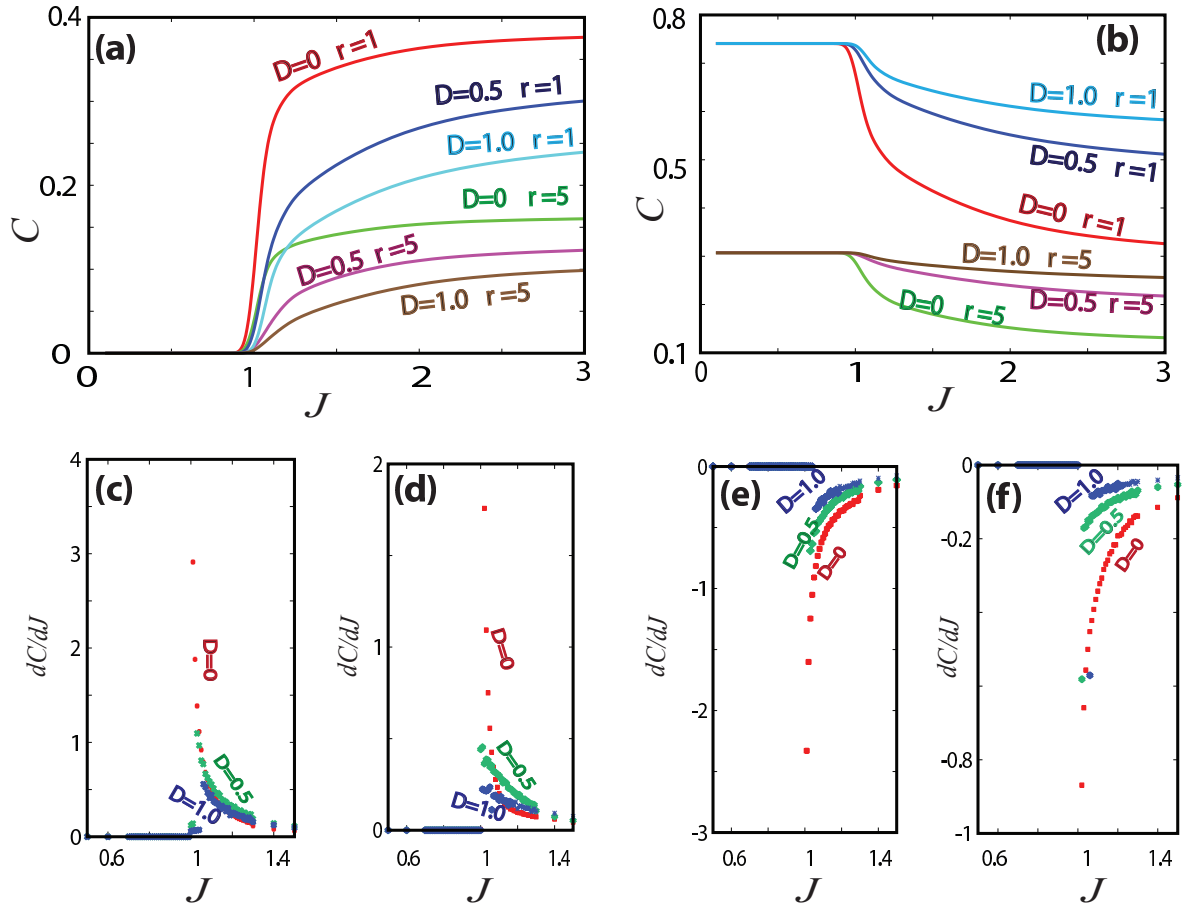}
\caption{The coherence $C$ of the isotropic XX model with DM interaction as a function of 
the spin-spin interaction strength $J$ is given in (a)(b)  for the $\sigma^z$-basis and 
the $\sigma^x$-basis respectively.  The coherence derivative $dC/dJ$ versus the spin-spin interaction strength
$J$ is given through in (c)(d).  (c)(d)  corresponds to 
$r=1$ and $r=5$ in the $\sigma^z$-basis and (e)(f) correspond to $r=1$ and $r=5$ in the 
$\sigma^x$-basis.}
\label{fig3}
\end{figure*}

Since the spin model shows singularities with respect to the derivative of the coherence the phase transition in this model is of second order.   Further we notice that the non-analytic behavior changes greatly with $D$, the DM coupling 
parameter. The non-analyticity is more  pronounced when $D=0$ and decreases with increase in the value of $D$.  The reason for this 
behavior being that, a higher value of $D$ means that the angle by which the spins are tilted are higher leading to a higher 
amount of the residual magnetism in the antiferromagnetic phase.  Thus the presence of DM interaction makes the phase transition less 
prominent in an XY-model.  Also we notice that the increase in distance has a direct bearing on the non-analytic behavior only when 
the DM interaction parameter is nonzero.  From Fig. \ref{fig2}(a)(b) we can see that the non-analytic nature of the 
XY model does not change with distance when $D=0$.  But when we change the antisymmetric exchange interaction 
coefficient to $D=0.5$ and $D=1$ the nature of the non-analytic behavior decreases with increase in distance.   
Throughout this study we find that the total coherence in the system is entirely due to the inter-qubit correlations referred to as 
intrinsic coherence.  Due to spin-flip symmetry there is no coherence which is localized within the qubit.

%
%
%
\subsection{Isotropic case: $\gamma=0$}
\label{xxdm}

The Heisenberg XY model with DM interaction discussed in the previous subsection is an anisotropic model.  
When the anisotropy parameter $\gamma=0$ the model becomes an isotropic. 

\begin{figure*}[ht]
\centering
\includegraphics[width=\linewidth]{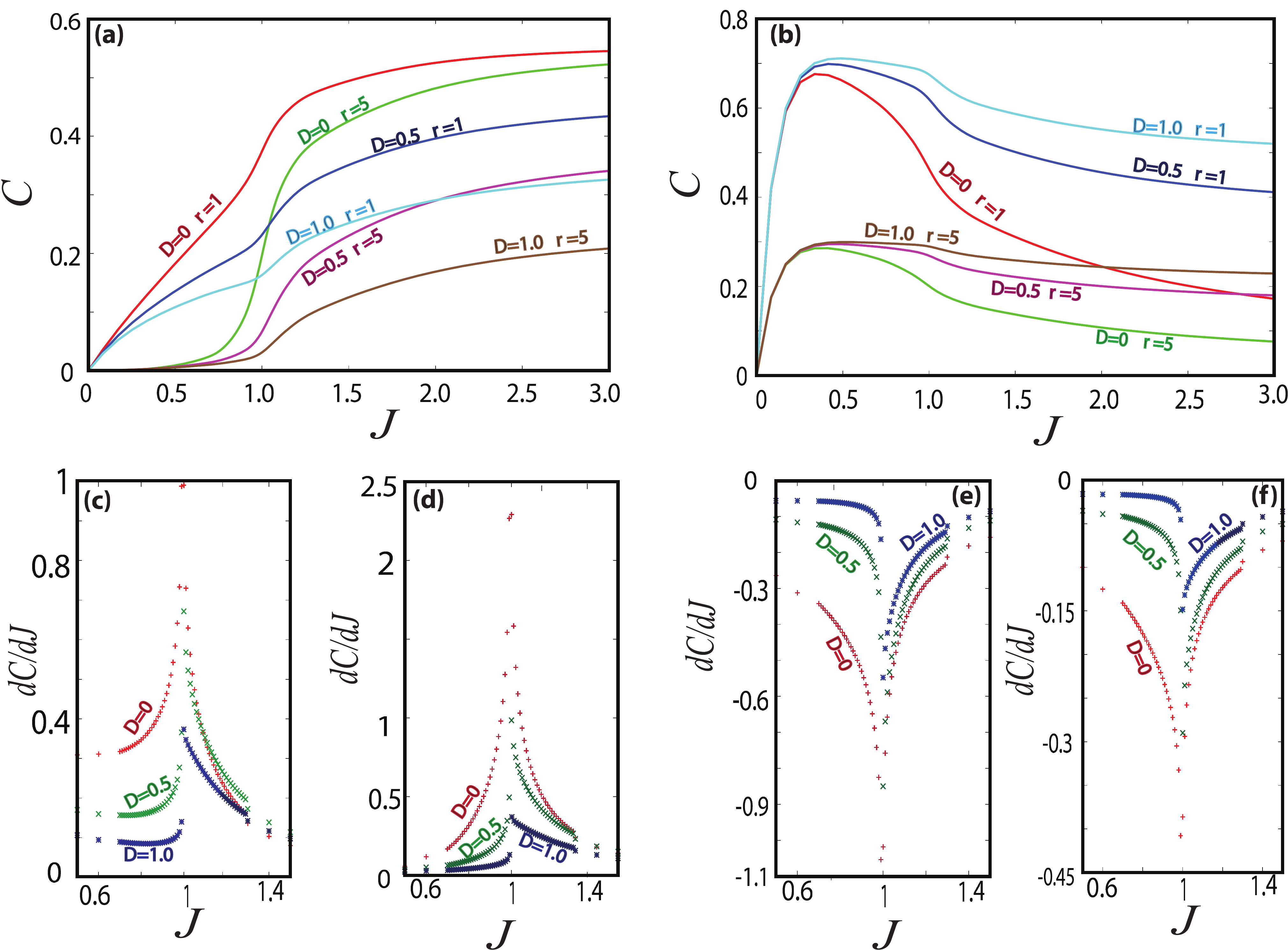}
\caption{The coherence of transverse Ising model $C$ with DM interaction as a function of 
the spin-spin interaction strength $J$ is given in (a)(b) for the $\sigma^z$-basis and 
the $\sigma^x$-basis respectively.  The coherence derivative $dC/dJ$ versus the spin-spin interaction strength
$J$ is given in (c)(d).  (c)(d)corresponds to 
$r=1$ and $r=5$ in the $\sigma^z$-basis and Fig (e)(f)correspond to $r=1$ and $r=5$ in the 
$\sigma^x$-basis.}
\label{fig4}
\end{figure*}

We observe from Fig. \ref{fig3}(a) that the coherence in the $\sigma^z$-basis is zero for low values of $J$ 
until we reach the critical point at $J=1$.  This is because the model exists in a unique ground state which is either 
$|\uparrow \uparrow ... \uparrow \rangle$ or $|\downarrow \downarrow ... \downarrow\rangle$ and 
consequently the density matrix is diagonal and hence there is no coherence in the system.  
When $J$ is greater than one, the antiferromagnetic state is a linear superposition of the many possible configurations which have 
zero magnetism.  Hence the density matrix is non-diagonal leading to the observation of coherence in the system.  The coherence 
does not show any discontinuity, so we investigate the derivative of coherence with respect to the spin-spin correlation function
$J$ in Fig. \ref{fig3}(c)(d). From the plots we observe that the first derivative of quantum coherence has a non-analytic behavior at 
$J=1$ which is the quantum critical point.  Further we observe that this non-analytic nature decreases with increase in the 
DM interaction parameter.  The $\sigma^x$-basis quantum coherence which is shown in Fig. \ref{fig3}(b) is not zero for 
the ferromagnetic phase, since the spin orientations are considered in the computational basis and the 
coherence is measured in the orthogonal $\sigma^x$-basis.  This leads to the situation where
the density matrix has off-diagonal elements resulting in a net coherence in the system.  
But the non-analyticity at the quantum critical points is still observed as shown in Fig.  \ref{fig3}(e)(f).

%
%
%
\subsection{Transverse Ising model with DM interaction: $\gamma=1$}
Another interesting model which can be obtained from the XY model given in  (\ref{xydmhamiltonian}) 
is the transverse field Ising model.  When the anisotropy factor is set to unity the  spin-spin 
interaction is only in the $\sigma^x$-direction and the field is in the $\sigma^z$-direction, that is orthogonal to the 
interaction direction.  Considering the case when $\gamma=1$ in (\ref{xydmhamiltonian})  we can arrive at the 
Hamiltonian of the transverse field Ising model with DM interaction where the $\sigma_{i}^{y} \sigma_{i+1}^{y}$
is removed. 

The functions (\ref{deltadefinition}) and (\ref{greensfunction}) reduce to the following relations
\begin{equation}
\Delta = \sqrt{[J(\cos \phi - 2D \sin \phi)-1]^{2} + J^{2}  \sin^{2} \phi},
\label{isingdeltadefinition}
\end{equation}
\begin{eqnarray}
G_{k} &=&  - \frac{2}{\pi} \int_{0}^{\pi} {\rm d} \phi \frac{\cos(\phi k)}{\Delta}
           [J (\cos \phi - 2 D \sin \phi) - 1] \nonumber \\ 
      &  &  + \frac{1}{\pi} 
            \int_{0}^{\pi} {\rm d}  \phi \frac{2 J \sin (\phi k)}{\Delta} \sin \phi, 
\label{isinggreensfunction}            
\end{eqnarray}
respectively.  Using these quantities in the magnetization (\ref{magnetizationxy}) and the two point 
correlation functions (\ref{xxcorrelator}), (\ref{yycorrelator}) and (\ref{zzcorrelator}) we can calculate 
the two qubit reduced density matrix.  From the knowledge of the two qubit density matrix 
we could compute the quantum coherence in the system using  
(\ref{coherencedef}) and the results are displayed in Fig.\ref{fig4}.

Figures \ref{fig4}(a)(b) show that the quantum coherence in the $\sigma^z$-basis is small in the ferromagnetic phase but increases after the 
quantum critical point ($J=1$) and saturates to a finite value whereas in the $\sigma^x$ basis it starts from zero then attains 
a maximum and then decreases and saturates to a finite value.  In the case of coherence in the $\sigma^z$-basis the saturation value
is higher when the DM parameter is lower and when the distance is also lower.  Meanwhile in the $\sigma^x$-basis, the coherence 
saturation value is lower when the antisymmetric exchange parameter $D$ is lower. As $J$ increases the 
contribution from the $\sigma^{x}$ and $\sigma^{y}$ terms increases and since they contribute to the off-diagonal 
elements they generate coherence in the $z$-basis.  Meanwhile in the $x$-basis the $\sigma^{x}$ terms contributes to 
the diagonal elements and this causes a decay of coherence in the system. 
 The quantum coherence does not show any 
discontinuity with respect to the spin-spin interaction parameter $J$.  In order to look for the order of the phase transition 
we plot the derivative of the quantum coherence with respect to $J$. From the plots in Figs \ref{fig4}(c)-(f) we find that the 
system is non-analytic at the point $J=1$ which denotes the quantum critical point. Hence we can again conclude that there is a second 
order phase transition in the model. The plots show that for the transverse Ising model the non-analytic nature 
decreases with respect to the DM interaction parameter and when measured in the $\sigma^z$-basis the non-analyticity increases with the 
distance between the spins and in the $\sigma^x$-basis the non-analytic behavior decreases with increase in distance between the 
spins.  

%
%
%
\section{Conclusions}
\label{conclusions}

The quantum phase transitions of the one dimensional XY-spin model with DM interaction was investigated in the 
thermodynamic limit at zero temperature, using the quantum coherence
in the system. The coherence in the models is calculated using a measure based on the quantum version of the Jensen-Shannon 
divergence which has both entropic and geometric properties. 
From our study we find that the behavior of quantum coherence                              
varies in the different phases with the change happening at the quantum critical point.  The quantum coherence itself does not 
show any discontinuity, but the first derivative of quantum coherence shows a non-analytic behavior at the $J=1$ point.    
Thus we conclude that the coherence detects the second order phase transition from the ferromagnetic phase to the 
antiferromagnetic phase.  Earlier works have used bipartite entanglement to study quantum phase
transitions in the model.  Due to the short ranged nature of bipartite entanglement, it gives an incomplete characterization of quantum phase 
transitions, which occur due to long-range correlations.  This was overcome by using multipartite entanglement measures to study 
them, which however is a computationally expensive quantity to calculate. Our results indicate that coherence could be a computationally simpler quantity to calculate that can robustly detect phase transitions.

The increase in the DM coupling decreases the non-analytic behavior near the quantum critical point.  This is because 
the two spins with the antisymmetric exchange coupling will not be exactly parallel to each other and makes a small angle 
between themselves which results in a weak ferromagnetic behavior even in the antiferromagnetic regime.  So the transition 
from the ferromagnetic phase to the antiferromagnetic phase is not as sharp as in a $D=0$ model 
i.e., without the spin-orbit coupling. This nature is also observed in the isotropic XX model and the transverse Ising model
with DM interaction.  Also for the XY model an increase in distance between the spins 
decreases the sharpness of the phase transition only when the DM interaction has a finite value.  In the case of the 
XX model and the Ising model this occurs irrespective of the presence of the DM interaction parameter $D$. 

From our results and past studies using other measures of coherence, it would appear that the coherence can be used as a tool to study quantum phase transitions.  Further QJSD measure of coherence distinguishes between the coherence which arises
due to the superposition between the levels (local coherence) and the coherence due to the correlations betweent the qubits 
(intrinsic coherence).
In the models that we have examined, it 
can detect the quantum critical point, the order of the phase transition and also captures the effects of 
DM coupling interactions accurately.  Through a more general extension of this study we can investigate one 
dimensional models with staggered DM interaction \cite{miyahara2007uniform,fisher1994random}.  It will also be worthwhile 
to study the role of coherence in discerning phase transitions in two dimensional spins models. Also studies on atom-photon 
models like the Jaynes-Cummings-Hubbard model \cite{greentree2006quantum,koch2009superfluid} will be an interesting topic 
for future research.  These studies will increase our understanding on the relationship between quantum coherence of a system 
and the quantum phase transition and will form the scope of our future work.

\begin{acknowledgements}
This work is supported by the Shanghai Research Challenge Fund, New York University Global Seed Grants for Collaborative Research, National Natural Science Foundation of China (Grant No. 61571301), the Thousand Talents Program for Distinguished Young Scholars (Grant No. D1210036A), and the NSFC Research Fund for International Young Scientists (Grant No. 11650110425), NYU-ECNU Institute of Physics at NYU Shanghai, Science and Technology Commission of Shanghai Municipality (Grant No. 17ZR1443600). IE would like to thank the Russian Science foundation (Grant No: 16-11-10218) for financial support. 
\end{acknowledgements}

%
%
%



%

\end{document}